\pdfoutput=1
\documentclass[smallextend]{svjour3}
\usepackage{mathptmx}
\usepackage{url}
\usepackage[hidelinks]{hyperref}
\usepackage[utf8]{inputenc}
\usepackage[T1]{fontenc}
\usepackage{caption}
\usepackage{xcolor}
\journalname{Applicable Algebra in Engineering, Communication and Computing}

\title{A Nonexistence Certificate for Projective Planes of Order Ten
with Weight 15 Codewords}
\titlerunning{A Nonexistence Certificate for Projective Planes of Order Ten
with Weight 15 Codewords}
\date{Received: October 30, 2019 / Accepted: February 14, 2020}

\author{Curtis Bright\and
Kevin Cheung\and
Brett Stevens\and
Dominique Roy\and
Ilias Kotsireas\and
Vijay Ganesh}
\institute{C.~Bright \at University of Waterloo \\ \email{cbright@uwaterloo.ca} \\ Webpage: \url{https://cs.uwaterloo.ca/~cbright/} \and
K.~Cheung, B.~Stevens \at Carleton University \and
D.~Roy \at Canada Revenue Agency \and
I.~Kotsireas \at Wilfrid Laurier University \and
V.~Ganesh \at University of Waterloo}
\authorrunning{Curtis Bright, Kevin Cheung, Brett Stevens, et al.}

\usepackage{multirow}
\usepackage{mathtools}
\usepackage{microtype}
\usepackage{amsmath}
\usepackage{amssymb}
\renewcommand{\~}{{\color{white}\textvisiblespace}}
\newcommand\elide{[\,\dots]}
\newcommand{\A}{\llap{\rotatebox{90}{>}}~}
\renewcommand{\v}{\llap{\rotatebox{90}{<}}~}
\newcommand{\macwilliams}{MacWilliams et al.~\shortcite{macwilliams1973existence}}
\let\shortcite\cite

\begin{document}

\maketitle

\begin{abstract}
Using techniques from the fields of symbolic computation and satisfiability checking
we verify one of the cases used in the landmark result that projective planes
of order ten do not exist.
In particular, we show that there exist no projective planes of order ten
that generate codewords of weight fifteen, a result
first shown in 1973 via an exhaustive computer search.
We provide a simple satisfiability (SAT) instance
and a certificate of unsatisfiability
that can be used to automatically verify this result
for the first time.  All previous 
demonstrations of this result have relied on search programs
that are difficult or impossible to verify---in fact, our search
found partial projective planes that were missed by previous searches
due to previously undiscovered bugs.
Furthermore, we show how the performance of the SAT solver
can be dramatically increased by employing functionality
from a computer algebra system (CAS).  Our SAT+CAS
search runs significantly faster than all
other published searches verifying this result.
\keywords{Combinatorial search \and Projective planes \and Symbolic computation \and Satisfiability checking \and SAT+CAS}
\end{abstract}

\section{Introduction}\label{sec:introduction}

A {projective plane} is a geometric structure
where parallel lines do not exist.  In other words, any two
lines in a projective plane must meet at some point, a property
that does not hold in the standard Euclidean plane.
The existence of non-Euclidean planes is initially counterintuitive
but they have been widely studied since the beginning of the 18th century.
As a simple example of this phenomenon, consider the case
of geometry on a sphere.  In this case the lines on a sphere
are the ``great circles'' of the sphere
and any two distinct lines intersect in exactly
two antipodal points.

A more exotic type of geometry known as \emph{finite geometry}
occurs when only a finite number of points exist.  In this article
we are concerned with finite projective geometry, i.e., geometry
that include axioms that say that
only a finite number of points
exist and that parallel lines do not exist.
A \emph{finite projective plane} is a model of the
finite projective geometry axioms (see Section~\ref{sec:preliminaries}
for the complete list).
In particular, a finite projective plane is said to be
of order $n$ if there are $n+1$ points on every line.

An important open question in finite geometry concerns the orders~$n$
for which finite planes exist.
Finite projective planes of order~$n$ can be constructed
whenever~$n$ is a prime power but it is unknown if any exist when~$n$
is not a prime power.  Despite a significant amount of effort
no one has ever been able to construct a finite plane in an
order~$n$ that is not a prime power and it has been widely conjectured
that such a plane cannot exist~\cite{bush1971unbalanced}.

A partial result was proven by Bruck and Ryser~\shortcite{bruck1949nonexistence}
who showed that $n$ must be the sum of two integer squares if a finite plane of
order $n$ exists with $n$ congruent to $1$ or $2\pmod{4}$.
Bruck and Ryser's result implies that a projective plane of order six
cannot exist.  Every other $n<10$ is a prime power and therefore
a finite plane of order $n$ does exist; the smallest
order that is not a prime power and not covered by the Bruck--Ryser
theorem is ten.

A first step towards solving the
existence question in order ten was completed
by MacWilliams, Sloane, and Thompson~\shortcite{macwilliams1973existence}.
In their paper the error-correcting code generated by a hypothetical projective
plane of order ten was studied.  In particular, they showed that the search could
be reduced to four cases that they called the weight $12$, $15$, $16$, and $19$
cases (see Section~\ref{sec:preliminaries}).  Furthermore, they used a computer search to show that the weight $15$ case
did not lead to a projective plane of order ten.

In the 1980s a number of extensive computer searches were performed to
settle the question of existence of a projective plane of order ten.
In particular, the weight $12$ case was solved by
Lam, Thiel, Swiercz, and McKay~\shortcite{lam1983nonexistence},
the weight $16$ case was solved
by Lam, Thiel, and Swiercz~\shortcite{lam1986nonexistence} (continuing work by
Carter~\shortcite{carter1974existence})
and the weight $19$ case was
solved by Lam, Thiel, and Swiercz~\shortcite{lam1989non}, finally showing that
a projective plane of order ten does not in fact exist.

Each of these cases required a significant amount of computational resources
to solve, including about~$2.7$ months of computing time on a CRAY-1A supercomputer
to solve the weight~$19$ case.  More recently, Roy~\shortcite{roy2011confirmation}
performed a verification of the nonexistence of the projective
plane of order ten using $3.2$ months of computing time with~$15$
CPU cores running at~$2.4$ GHz.
Several recent works~\cite{casiello2010sull,clarkson2014nonexistence,perrott2016existence}
have also performed verifications of the weight~$15$ case using custom-written search code.
These verifications used the programming language C
or the programming languages of the computer algebra systems GAP and Mathematica.
Additionally, Bruen and Fisher~\shortcite{bruen1973blocking} showed that
the weight~$15$ case can be solved using a result of Denniston~\shortcite{denniston1969non}
but this was also obtained via a computer search.

In this paper we perform a verification of the weight~$15$ case
using the properties derived by MacWilliams, Sloane, and Thompson~\shortcite{macwilliams1973existence}.
Our verification is unique
in that we translate the properties of a projective plane into Boolean
logic and then perform the search using a SAT solver.
SAT solvers are known to be some of the best tools to perform
combinatorial searches; for example, 
Heule, Kullmann, and Marek~\shortcite{heule2017solving}
state that today they are the ``best solution''
for most kinds of combinatorial searches.
Even so, they mention that there are some problems that
SAT solvers have not yet been successfully applied to.
In fact, they explicitly list the 
search for a projective plane of order ten as one of these problems:
\begin{quote}
An example where only a solution by [special purpose solvers]
is known is the determination that there is no projective plane of order 10 \elide\
To the best of our knowledge the effort has not been replicated, and
there is definitely no formal proof.
\end{quote}

The fact that we perform our search using a SAT solver
means that we can
produce a formally verifiable \emph{certificate} that the weight~$15$ case does not lead to
a solution.  In contrast to all previous searches that have been completed,
one can verify our results
without needing to trust the particular choice of hardware, compiler,
or search algorithms that we happened to use in our verification.
Instead, one merely needs to trust our encoding of the problem into SAT
(see Section~\ref{sec:encoding})
and our SAT instance generation script (see Section~\ref{sec:results}).

We do not claim our verification is a \emph{formal proof} of nonexistence
because it relies on mathematical results that (at least currently)
have no machine-verifiable formal proof.
However, compared to previous approaches our verification has the advantage that
it is not necessary to trust
code that implements a search algorithm.
This is particularly important considering that efficient search algorithms
often need to be written in a convoluted way to obtain optimum performance.
In fact, while verifying that our SAT encoding
was producing correct results we uncovered bugs in previous searches
(see Section~\ref{sec:results}).

Our result is also a first step towards a formal proof.
The SAT encoding is deliberately chosen to be as simple as possible so that
(1) the possibility of an encoding bug is less likely, and
(2) it will be as simple as possible to formally generate the SAT clauses
directly from the axioms that define a projective plane of order ten.
This approach of reducing a problem to SAT, solving the resulting SAT
instance, formally verifying the nonexistence certificate, and then
finally formally verifying the SAT encoding in a theorem prover
has recently successfully formally verified
the proofs of the Boolean Pythagorean triples conjecture~\cite{cruz2018formally,heule2016solving}
and the Erd\H{o}s discrepancy conjecture for discrepancy up to three~\cite{keller2019,konev2015computer}.

We also show how to use a computer algebra system (CAS) to greatly improve
the efficiency of the SAT solver (see Section~\ref{sec:symbreaking}).
This ``SAT+CAS'' approach of combining SAT solvers
with computer algebra systems has recently been applied to a variety of problems
including verifying the correctness of Boolean arithmetic circuits~\cite{kaufmann2019verifying},
finding new algorithms for $3\times3$ matrix multiplication~\cite{heule2019new},
and finding or disproving the existence of certain kinds of combinatorial designs~\cite{bright2019good}.
For more detailed surveys on the SAT+CAS paradigm and the kinds of problems that it has been applied
to see~\cite{bright2019sat,davenport2019symbolic}.
Our SAT+CAS approach for solving the weight 15 case performs better than
all other searches that have been
previously published (see Section~\ref{sec:comparison}).

\section{Projective plane preliminaries}\label{sec:preliminaries}

A finite projective plane of order $n$ consists of a set of lines and 
a set of points that satisfy the following axioms:
\begin{description}
\item[P1.] There are $n+1$ points on every line and there are $n+1$ lines through every point.
\item[P2.] There is exactly one line between every two distinct points and
every two distinct lines intersect in exactly one point.
\end{description}
A consequence of these axioms is that a finite projective plane
of order~$n$ contains exactly $n^2+n+1$ points and exactly $n^2+n+1$ lines~\cite{kaahrstrom2002projective}.

One convenient way of representing a finite projective plane is by an incidence matrix
that encodes which points lie on which lines. This matrix has a~$1$
in the $(i,j)$th entry if point~$j$ is on line~$i$ and has a~$0$ in the $(i,j)$th entry otherwise.
In this representation axiom~P2 says that every pair of columns
or pair of rows intersect exactly once (where two columns or rows
\emph{intersect} if they both contain a~$1$ in the same location).
The number of times that two columns or rows intersect is given by the inner
product (over the reals) of the two columns or rows, so axiom~P2 says that the inner product
product of any two columns or rows is exactly one.

It follows that a finite projective plane of order ten is equivalent
to a $\{0,1\}$-matrix of size $111\times111$ such that:
\begin{description}
\item[P1.] Every row and column contains exactly eleven $1$s.
\item[P2.] The inner product of any two distinct rows or two distinct columns
is exactly $1$.
\end{description}
The ordering of the rows and columns of the incidence matrix
of a projective plane is arbitrary and we say that two matrices
are \emph{equivalent} if one matrix can be transformed into the
other by a reordering of the rows and columns.

Suppose $P$ is the incidence matrix of a hypothetical
projective plane of order ten.
The elements of the row space of~$P$ (mod~$2$) are known as the
\emph{codewords} of~$P$ and the number of~$1$s in a codeword is known
as the \emph{weight} of the codeword.
Let $w_k$ denote the number of codewords of $P$ of weight $k$.
For example,
$w_0=1$ because the zero vector is in the row space of $P$
and no other codeword has a weight of zero.
Also, it was shown by Assmus and Mattson~\shortcite{assmus1970possibility}
that the values of all $w_k$ for $0\leq k\leq 111$ can be determined
from just the values of $w_{12}$, $w_{15}$, and $w_{16}$.

The relationships between the values of $w_k$ are what ultimately
lead to the contradiction that showed that a projective
plane of order ten cannot exist.
For example,
\[ w_{12}=w_{15}=w_{16}=0 \qquad \text{imply that} \qquad w_{19}=24{,}675. \]
However, a number of 
exhaustive computer searches~\cite{carter1974existence,%
lam1986nonexistence,lam1989non,lam1983nonexistence,macwilliams1973existence}
found no codewords of weights $12$, $15$, $16$, or $19$---%
thus implying the hypothetical projective plane~$P$ cannot exist.

In fact, it is known that
$w_{15}$ and $w_{19}$ cannot both be zero~\shortcite{carter1974existence,hall1980configurations}.
Furthermore, Carter~\cite{carter1974existence} and Hall~\shortcite{hall1980configurations}
show that the weight 19 codewords either
arise from weight 12 codewords, weight 16 codewords, or are of a third kind that
they call primitive.  They show that if $w_{15}$ and $w_{16}$ are zero
then the number of primitive weight 19 codewords must be positive.
Thus, to show that a projective plane of order ten does not
exist it suffices to show that there exists no codewords of
$P$ of weights $15$, $16$, or primitive weight~$19$ codewords.
In the remainder of this
paper we describe the construction of a SAT instance that
necessarily has a solution if a codeword of~$P$ of weight~$15$
exists.  We then provide a certificate that this SAT
instance is unsatisfiable and therefore solve one of the
three cases necessary to prove the nonexistence of a
projective plane of order ten.

\subsection{Incidence matrix structure}\label{subsec:structure}

\macwilliams{} derive a number of properties that the
structure of a projective plane of order ten
with weight 15 codewords must satisfy.  In particular, they show
that up to equivalence the incidence matrix of such a projective
plane can be partitioned into a $3\times3$ grid of submatrices as
follows:
\[
\begin{matrix}
& & 15 & ~60~ & 36 \\
\llap{\text{heavy~~~~~~}} \phantom{0}6~ & \multirow{3}{*}{{$\left(\rule{0pt}{16pt}\right.$}} & 6 & 0 & 5 & \multirow{3}{*}{{$\left.\rule{0pt}{16pt}\right)$}} \\
\llap{\text{medium~~~~~~}} 15~ & & 3 & 8 & 0 \\
\llap{\text{light~~~~~~}} 90~ & & 1 & 6 & 4
\end{matrix}
\]
Here the numbers outside the matrix denote the number
of rows or columns in that part of the matrix and the numbers
inside the matrix are the number of $1$s that appear
in each row of the submatrix in that part of the matrix.
\macwilliams{} call the first 15 columns the $A$ points,
the next 60 columns the $C$ points, and the remaining columns
the $B$ points.  Furthermore, Roy~\cite{roy2011confirmation}
calls the first 6 rows the \emph{heavy} lines, the next 15 rows
the \emph{medium} lines, and the remaining rows the \emph{light} lines.

\macwilliams{} also
show that up to equivalence there is exactly one way
of assigning the $1$s in each submatrix except for the last
two submatrices of the last row.  Furthermore they provide an explicit
representation of the unique assignment up to equivalence.

\subsection{Initial entries}\label{subsec:initial}

Up to equivalence, a number of entries of a projective plane
of order ten containing a weight~$15$ codeword
can be initialized in advance, including all entries in the first~$21$ rows
and~$15$ columns~\cite{macwilliams1973existence}.
In our search we focus on the first $75$ columns and
$51$ rows of the projective plane.  The entries of this submatrix that we
fix in advance are shown in Figure~\ref{fig:matrix}.

\begin{figure*}
\centering\small\tt
\mbox{~~~~~~<~points 1-15~><~~~~~~~~~~~~~~~~~~~~~~~points 16-75~~~~~~~~~~~~~~~~~~~~~~~>}\\\mbox{~~~~~}\\[-0.5\baselineskip]
\mbox{~~~~~\A111110000000000000000000000000000000000000000000000000000000000000000000000}
\mbox{~~~~~~100001111000000000000000000000000000000000000000000000000000000000000000000}
\mbox{~~~~~\llap{lines}~010001000111000000000000000000000000000000000000000000000000000000000000000}
\mbox{~~~~~\llap{1-6 }~001000100100110000000000000000000000000000000000000000000000000000000000000}
\mbox{~~~~~~000100010010101000000000000000000000000000000000000000000000000000000000000}
\mbox{~~~~~\v000010001001011000000000000000000000000000000000000000000000000000000000000}
\mbox{~~~~~\A100000000100001000000000000000000000001111110000000010000100000000000000000}
\mbox{~~~~~~100000000010010000000111111000000000000000000000000000100000010000000000000}
\mbox{~~~~~~100000000001100000000000000111111000000000000000100000000000000000100000000}
\mbox{~~~~~~010000100000001111111000100000000000000000000000000000000000000000100000000}
\mbox{~~~~~~010000010000010000000000000000010001000000000000000000000101100010000100001}
\mbox{~~~~~~010000001000100000000000010000000000010000010110000000000010000100000001000}
\mbox{~~~~~~001001000000001000000001000000000010000000000000111000001000000110000000000}
\mbox{~~~~~\llap{lines}~001000010001000000100000001000000000000100000001000100000000000000000001110}
\mbox{~~~~~\llap{7-21}~001000001010000010000000000000001000000001000000000001000000001000011000001}
\mbox{~~~~~~000101000000010001000000000100000000000000001100000111010000000000000000000}
\mbox{~~~~~~000100100001000000000100000000000111111000000000000000000000000000001000000}
\mbox{~~~~~~000100001100000000001000000000100000000000000000000000101000000001000110100}
\mbox{~~~~~~000011000000100100000010000000000000100010000001000000000000100000010010000}
\mbox{~~~~~~000010100010000000000000000010000000000000101000010000000011000001000000010}
\mbox{~~~~~\v000010010100000000010000000001000100000000000010001000010000011000000000000}
\mbox{~~~~~\A1000000000000001\~\~\~\~\~000000000000\~\~\~\~\~000000\~\~\~\~0\~\~\~0\~0\~\~0\~\~\~0\~\~\~\~0\~\~\~\~\~\~\~\~}
\mbox{~~~~~~100000000000000\~1\~\~\~\~000000000000\~\~\~\~\~000000\~\~\~\~0\~\~\~0\~0\~\~0\~\~\~0\~\~\~\~0\~\~\~\~\~\~\~\~}
\mbox{~~~~~~100000000000000\~\~1\~\~\~000000000000\~\~\~\~\~000000\~\~\~\~0\~\~\~0\~0\~\~0\~\~\~0\~\~\~\~0\~\~\~\~\~\~\~\~}
\mbox{~~~~~~100000000000000\~\~\~1\~\~000000000000\~\~\~\~\~000000\~\~\~\~0\~\~\~0\~0\~\~0\~\~\~0\~\~\~\~0\~\~\~\~\~\~\~\~}
\mbox{~~~~~~100000000000000\~\~\~\~1\~000000000000\~\~\~\~\~000000\~\~\~\~0\~\~\~0\~0\~\~0\~\~\~0\~\~\~\~0\~\~\~\~\~\~\~\~}
\mbox{~~~~~~100000000000000\~\~\~\~\~1000000000000\~\~\~\~\~000000\~\~\~\~0\~\~\~0\~0\~\~0\~\~\~0\~\~\~\~0\~\~\~\~\~\~\~\~}
\mbox{~~~~~~000000000100000\~\~\~\~001\~\~\~\~\~\~\~00\~\~0\~\~\~\~000000\~\~0\~\~\~0\~0\~0000\~\~\~00\~\~0\~\~\~00\~0\~\~}
\mbox{~~~~~~000000000100000\~\~\~\~00\~1\~\~\~\~\~\~00\~\~0\~\~\~\~000000\~\~0\~\~\~0\~0\~0000\~\~\~00\~\~0\~\~\~00\~0\~\~}
\mbox{~~~~~~000000000100000\~\~\~\~00\~\~1\~\~\~\~\~00\~\~0\~\~\~\~000000\~\~0\~\~\~0\~0\~0000\~\~\~00\~\~0\~\~\~00\~0\~\~}
\mbox{~~~~~~000000000100000\~\~\~\~00\~\~\~1\~\~\~\~00\~\~0\~\~\~\~000000\~\~0\~\~\~0\~0\~0000\~\~\~00\~\~0\~\~\~00\~0\~\~}
\mbox{~~~~~~000000000100000\~\~\~\~00\~\~\~\~1\~\~\~00\~\~0\~\~\~\~000000\~\~0\~\~\~0\~0\~0000\~\~\~00\~\~0\~\~\~00\~0\~\~}
\mbox{~~~~~~000000000100000\~\~\~\~00\~\~\~\~\~1\~\~00\~\~0\~\~\~\~000000\~\~0\~\~\~0\~0\~0000\~\~\~00\~\~0\~\~\~00\~0\~\~}
\mbox{~~~~~~000000000000001000000\~\~00\~\~1\~\~\~\~\~\~0\~\~\~000000\~\~\~\~000\~0\~\~\~00\~\~\~\~\~00\~0\~\~\~\~\~\~\~\~}
\mbox{~~~~~~000000000000001000000\~\~00\~\~\~1\~\~\~\~\~0\~\~\~000000\~\~\~\~000\~0\~\~\~00\~\~\~\~\~00\~0\~\~\~\~\~\~\~\~}
\mbox{~~~~~\llap{lines}~000000000000001000000\~\~00\~\~\~\~1\~\~\~\~0\~\~\~000000\~\~\~\~000\~0\~\~\~00\~\~\~\~\~00\~0\~\~\~\~\~\~\~\~}
\mbox{~~~~~\llap{22-51}~000000000000001000000\~\~00\~\~\~\~\~1\~\~\~0\~\~\~000000\~\~\~\~000\~0\~\~\~00\~\~\~\~\~00\~0\~\~\~\~\~\~\~\~}
\mbox{~~~~~~000000000000001000000\~\~00\~\~\~\~\~\~1\~\~0\~\~\~000000\~\~\~\~000\~0\~\~\~00\~\~\~\~\~00\~0\~\~\~\~\~\~\~\~}
\mbox{~~~~~~000000000000001000000\~\~00\~\~\~\~\~\~\~1\~0\~\~\~000000\~\~\~\~000\~0\~\~\~00\~\~\~\~\~00\~0\~\~\~\~\~\~\~\~}
\mbox{~~~~~~000000000010000\~0\~\~\~\~000000\~0\~\~\~01\~\~\~\~\~\~\~00\~0\~\~\~\~0\~\~\~00\~\~\~00\~00\~\~0\~00\~\~\~\~00}
\mbox{~~~~~~000000000010000\~0\~\~\~\~000000\~0\~\~\~0\~1\~\~\~\~\~\~00\~0\~\~\~\~0\~\~\~00\~\~\~00\~00\~\~0\~00\~\~\~\~00}
\mbox{~~~~~~000000000010000\~0\~\~\~\~000000\~0\~\~\~0\~\~1\~\~\~\~\~00\~0\~\~\~\~0\~\~\~00\~\~\~00\~00\~\~0\~00\~\~\~\~00}
\mbox{~~~~~~000000000010000\~0\~\~\~\~000000\~0\~\~\~0\~\~\~1\~\~\~\~00\~0\~\~\~\~0\~\~\~00\~\~\~00\~00\~\~0\~00\~\~\~\~00}
\mbox{~~~~~~000000000010000\~0\~\~\~\~000000\~0\~\~\~0\~\~\~\~1\~\~\~00\~0\~\~\~\~0\~\~\~00\~\~\~00\~00\~\~0\~00\~\~\~\~00}
\mbox{~~~~~~000000000010000\~0\~\~\~\~000000\~0\~\~\~0\~\~\~\~\~1\~\~00\~0\~\~\~\~0\~\~\~00\~\~\~00\~00\~\~0\~00\~\~\~\~00}
\mbox{~~~~~~000000000000010\~\~0\~\~\~0000000\~\~\~0\~\~\~0\~\~1\~\~\~\~\~00\~\~\~\~\~00000\~0\~000\~\~0\~\~\~\~0\~\~\~\~0}
\mbox{~~~~~~000000000000010\~\~0\~\~\~0000000\~\~\~0\~\~\~0\~\~\~1\~\~\~\~00\~\~\~\~\~00000\~0\~000\~\~0\~\~\~\~0\~\~\~\~0}
\mbox{~~~~~~000000000000010\~\~0\~\~\~0000000\~\~\~0\~\~\~0\~\~\~\~1\~\~\~00\~\~\~\~\~00000\~0\~000\~\~0\~\~\~\~0\~\~\~\~0}
\mbox{~~~~~~000000000000010\~\~0\~\~\~0000000\~\~\~0\~\~\~0\~\~\~\~\~1\~\~00\~\~\~\~\~00000\~0\~000\~\~0\~\~\~\~0\~\~\~\~0}
\mbox{~~~~~~000000000000010\~\~0\~\~\~0000000\~\~\~0\~\~\~0\~\~\~\~\~\~1\~00\~\~\~\~\~00000\~0\~000\~\~0\~\~\~\~0\~\~\~\~0}
\mbox{~~~~~\v000000000000010\~\~0\~\~\~0000000\~\~\~0\~\~\~0\~\~\~\~\~\~\~100\~\~\~\~\~00000\~0\~000\~\~0\~\~\~\~0\~\~\~\~0}
\caption{Our initial instantiation of the first $75$ columns and $51$ rows of $P$ where blank spaces represent unknown entries.}\label{fig:matrix}
\end{figure*}

The first 6 rows (the heavy lines) of the projective plane are taken to be
identical with the representation of \macwilliams{}.  The next 21 rows (the medium
lines) are equivalent to MacWilliams' representation---we have only
applied a column permutation to their representation to more clearly explain how
we assigned the $1$s in the later rows.  In particular, the specific ordering of
columns 16--75 that was used in Figure~\ref{fig:matrix} was chosen in order to allow initializing
the diagonal line of $1$s that appears in rows 22--51 (see below).

The first $15$ columns in Figure~\ref{fig:matrix}
are also specified to be identical to the representation given by \macwilliams{}.
They choose to order the rows so that the light rows
containing a~$1$ in the first column appear first, followed by the light
lines containing a~$1$ in the tenth, fifteenth, eleventh, and fourteenth columns
(in that order).  This ordering was chosen in an attempt to maximize
the number of overlapping columns with unassigned entries in the light rows.
While searching for completions of the unassigned entries, conflicts between two
light rows will occur in columns where both light rows
contain unassigned entries---therefore maximizing the number of overlapping
columns with unassigned entries tends to increase the number of conflicts
and speed up the search.

This leaves the lower-right $30\times60$ submatrix of
Figure~\ref{fig:matrix}.  The zeros that appear in
this submatrix are easily determined; if they were~$1$s
then the column that they are on would intersect more than
once with another column.
For example, consider the 22nd entry of the 22nd line---if this
entry was a~$1$ then the first and 22nd column would
intersect twice, in contradiction
to the matrix of Figure~\ref{fig:matrix} being
a partial projective plane.

Next we show that the diagonal line of $1$s that appears
on the left of the $30\times60$ submatrix can be assumed
without loss of generality.
For example, consider the light rows that contain a~$1$
in the first column (rows $22$--$27$).
By the projective plane axiom P2,
these lines must share a point of intersection with the fourth medium
line (row~$10$).  By inspection, there are exactly six possible columns
for this point of intersection (namely, columns $16$--$21$).
In other words, there must be a~$1$ in each row of the submatrix given
by rows $22$--$27$ and columns $16$--$21$.

Since each of the rows $22$--$27$ are already pairwise intersecting
in the first column they must not be pairwise intersecting
in the columns $16$--$21$.  Similarly, each of the columns $16$--$21$
are already pairwise intersecting in the tenth row so they must not
be pairwise intersecting in the rows $22$--$27$.
In other words, each row and column of the submatrix given
by rows $22$--$27$ and columns $16$--$21$ contains at most a single~$1$.
By reordering the rows $22$--$27$ we can assume
without loss of generality that the submatrix given
by rows $22$--$27$ and columns $16$--$21$ is the identity matrix.

This explains why we may initialize the diagonal $1$s that appear in the
rows $22$--$27$ of Figure~\ref{fig:matrix}.  The same reasoning
explains the initializations in the rows $28$--$33$ (with columns $22$--$27$), rows $34$--$39$
(with columns $28$--$33$), rows $40$--$45$ (with columns $34$--$39$),
and rows $46$--$51$ (with columns $39$--$44$).  Note that in the final case
the set of columns that are used overlaps with columns that are used in the second last case,
and this causes the diagonal to be offset in the last six rows.

Once the $1$s in the lower-right submatrix
of Figure~\ref{fig:matrix} have been assigned some previously undetermined
entries can be set to $0$ but for simplicity we do not include
these in Figure~\ref{fig:matrix}.
In any case, it is mostly inconsequential if these entries
are included in our initial instantiation or not.
If they are not given the SAT solver will almost immediately discern these entries
using Boolean constraint propagation on the clauses used in our
projective plane encoding.

The entries of Figure~\ref{fig:matrix} were all derived using mathematical arguments
and not via a computer search.
A SAT solver could also be used to derive some of these entries,
but this would require a more complicated encoding (see Section~\ref{sec:encoding}).
Thus, we prefer to take
the entries of Figure~\ref{fig:matrix} as given and fixed in advance.

\section{SAT encoding}\label{sec:encoding}

In this section we describe the encoding we used to show the nonexistence of codewords of weight~15
in a projective plane of order ten.
Our encoding uses the Boolean variables $p_{i,j}$ where $i$ and~$j$
are between $1$ and $111$.
When $p_{i,j}$ is true it represents
that the $(i,j)$th entry of $P$ is $1$ and when $p_{i,j}$ is false
it represents that the $(i,j)$th entry of $P$ is $0$.
Thus, when the $(i,j)$th entry in Figure~\ref{fig:matrix} is a~$1$
we include the unit clause $p_{i,j}$ in our SAT instance and
when the entry is a~$0$ we include the unit clause $\lnot p_{i,j}$
in our SAT instance.

\subsection{Incidence constraints}

We now describe the constraints we used to specify that the
incidence matrix defined by the Boolean variables $p_{i,j}$
forms a projective plane.  In particular, axiom~P2 from Section~\ref{sec:preliminaries}
says that all rows and columns intersect exactly once.
We encode this axiom by splitting it up into the following two constraints:
\begin{enumerate}
\item The pairwise row and column inner products of~$P$ are \emph{at most} one.
\item The pairwise row and column inner products of~$P$ are \emph{at least} one.
\end{enumerate}
Furthermore, in the second case, we found that it was only necessary to consider
inner products between the medium rows and the light rows and the inner products
between the first 15 columns and the later columns.  We also only used the first
51 rows and 75 columns of~$P$.  Our searches found no satisfying assignments of even this
strictly smaller set of constraints.
The fact that these constraints are unsatisfiable
therefore shows more than just the nonexistence of weight 15 codewords; we also show the nonexistence
of partial projective planes that complete Figure~\ref{fig:matrix}.

We do not directly encode axiom~P1 in our SAT instances.  This axiom is
present merely to exclude ``degenerate'' cases from being considered as projective
planes
(where the rows of degenerate projective planes of order ten
have weights $1$, $2$, $110$, or $111$).
However, an examination of even the first
row of Figure~\ref{fig:matrix} shows that degenerate cases are naturally
excluded by our encoding.  Thus, the completions of $P$ that satisfy axiom~P2
naturally satisfy axiom~P1.  It is possible to encode axiom~P1 in conjunctive
normal form (for example, by using a sequential counter encoding~\cite{sinz2005towards}).  However, this
introduces new variables and in our experiments decreased the performance
of the SAT solver.

\subsubsection{Encoding that columns and rows intersect at most once}\label{subsec:nonintersection}

Consider rows $i$ and $j$ for arbitrary $1\leq i,j\leq111$ with $i\neq j$.
To enforce that these rows intersect at most once we must enforce that
there do not exist column indices $k$ and $l$ (where $1\leq k,l\leq111$ and $k\neq l$)
such that
$p_{i,k}$, $p_{i,l}$, $p_{j,k}$, and $p_{j,l}$ are all simultaneously true.
In other words, for each pair of distinct indices $(i,j)$ and $(k,l)$ at least one
variable $p_{i,k}$, $p_{i,l}$, $p_{j,k}$, or $p_{j,l}$ must be false; this
also implies that pairwise all columns intersect at most once.
As clauses in conjunctive normal form we encode this as
\[ \bigwedge_{i<j}\bigwedge_{k<l}(\lnot p_{i,k}\lor\lnot p_{i,l}\lor\lnot p_{j,k}\lor\lnot p_{j,l}) . \]

\subsubsection{Encoding that columns and rows intersect at least once}\label{subsec:intersection}

Consider rows $i$ and $j$ for arbitrary $1\leq i,j\leq111$ with $i\neq j$.
To enforce that these rows intersect at least once we must enforce that
there exist a column index $k$ with $1\leq k\leq 111$ such that $p_{i,k}$
and $p_{j,k}$ are simultaneously true.  This can be encoded as
\[ \bigwedge_{i<j}\bigvee_k(p_{i,k}\land p_{j,k}), \]
however, this formula is not in conjunctive normal form (CNF) and therefore cannot be used
directly with a standard SAT solver.  However, in certain cases this formula
easily simplifies into a formula in CNF.

In particular, consider the case when $i$ is the index
of a medium line and $j$ is the index of a light line (i.e., $7\leq i\leq 21$ and $22\leq j\leq 111$).
In this case, the truth values of the variables $p_{i,k}$ for all $1\leq k\leq 111$ can be determined in advance.
Their values are forced by the unit clauses encoding the entries displayed in Figure~\ref{fig:matrix}
and the fact that each row~$i$ is already known to contain eleven~$1$s (so $p_{i,k}$
must be false for $76\leq k\leq 111$).

Let $S(i)$ denote the indices $k$ such that $p_{i,k}$ is true.  Then
\[ \bigvee_{1\leq k\leq111}(p_{i,k}\land p_{j,k}) \quad\text{simplifies to}\quad \bigvee_{k\in S(i)}p_{j,k} . \]
Furthermore, the expression on the right can be determined in advance
since the entries of row~$i$ of~$P$ are completely known.

We encode the fact that the columns $k$ and $l$ intersect at least once
(where $1\leq k\leq 15$ and $16\leq l\leq 75$) in a similar way.  Let $T(k)$ denote the set of row indices~$i$
such that $p_{i,k}$ is true; since the first 15 columns of~$P$ are known these sets
can be determined in advance.  Then we encode the fact that columns $k$ and $l$ intersect at least once
by \( \bigvee_{i\in T(k)}p_{i,l} . \)

Altogether we encode these constraints in conjunctive normal form by
\[ \bigwedge_{\substack{7\leq i\leq 21\\22\leq j\leq 51}}\,\bigvee_{k\in S(i)}p_{j,k} \qquad\text{and}\qquad \bigwedge_{\substack{16\leq l\leq75\\k\in\{1,10,11,14,15\}}}\,\bigvee_{i\in T(k)}p_{i,l} . \]
Note that we have limited ourselves to only using variables from the first 51 rows and 75 columns
in these clauses---in the right formula we only use the $k$ for which the entries in $T(k)$ are
from the set $\{22,\dotsc,51\}$.

More than 51 rows and 75 columns can be used but these values
were chosen in an attempt to minimize the number of variables
necessary to show the impossibility completing the initial values of $P$
into a complete projective plane.
In particular, the entries of $S(i)$ are from the set $\{16,\dotsc,75\}$ so
it is necessary to use at least 75 columns.  Using a smaller number of rows
is possible---for example, only 45 rows could be used by taking $k\in\{1,10,11,15\}$ in the right formula.
However, this SAT instance was experimentally
found to be satisfiable (see Figure~\ref{fig:partial} in Section~\ref{sec:implementation}).

\subsection{Symmetry breaking}\label{sec:symbreaking}

When search spaces are highly symmetric, SAT solvers generally perform poorly
because they typically have not been optimized to detect symmetries.  Thus, in the
presence of symmetry a SAT solver will tend to repeat the same search for every symmetry
that exists.
A common method of improving their performance is to add constraints
that eliminate or ``break'' the symmetry~\cite{crawford1996symmetry}.
Symmetry breaking is not essential to our method, but it does greatly increase
its effectiveness.  In Section~\ref{sec:results}
we provide timings for our method both with and without symmetry breaking.

The \emph{symmetry group} of a given matrix is the set of row and column permutations that fix its entries.
As an explicit example, consider the upper-left $6\times15$ submatrix of~$P$:
\begin{center}\tt
111110000000000 \\
100001111000000 \\
010001000111000 \\
001000100100110 \\
000100010010101 \\
000010001001011
\end{center}
The symmetry group of this matrix is isomorphic to $S_6$, the symmetric group of degree~$6$.
It contains $6!=720$ distinct permutations, including, for example, the
permutation that swaps the first two rows and swaps column $i$ with column $i+4$
for $2\leq i\leq5$.
The symmetry group of the
upper-left $21\times75$ submatrix of~$P$ is also isomorphic to $S_6$ and we call this symmetry group~$S$.

We may apply the permutations from~$S$ to partial completions
of the first 75 columns of~$P$.
Such an action necessarily produces another equivalent partial completion
(up to a reordering of the light rows).
In our search, we attempt to eliminate as many
equivalent partial completions from the search space as possible,
leaving only partial completions that are not equivalent to each other.

We now focus on the first six light rows (rows $22$--$27$) and the completions of those rows.
We use the permutations in~$S$ to transform completions of
this submatrix into other equivalent completions of this submatrix.
We only consider permutations of~$S$ that fix the first column of this submatrix,
since it is not possible for any permutation that moves the first column to fix
the upper-left $27\times75$ submatrix of Figure~\ref{fig:matrix} (due to the $1$s on the first
column of rows $22$--$27$).
The subgroup of~$S$ fixing the first column of the upper-left $21\times75$
submatrix of $P$ is isomorphic to $S_4\times S_2$
and contains $4!\cdot2=48$ permutations.

\subsubsection{Programmatic SAT}\label{sec:programmatic}

We now describe our symmetry breaking method using the programmatic SAT paradigm~\cite{ganesh2012lynx}.
A \emph{programmatic SAT} solver can learn clauses in a programmatic fashion, for example through a piece of code that
queries a computer algebra system.
In our application, whenever a completion of the upper-left $27\times75$ submatrix of~$P$ is found
we record it and programmatically learn clauses that block the completion as well as all completions that are
equivalent to it.  The search then continues until all inequivalent completions have been recorded.

See Figure~\ref{fig:completion} for an explicit example of such a completion.
Say that $C$
is the set of variables whose values were initially unknown but have been assigned to true
in a completion.  The set~$C$ must contain exactly 30 variables since there are 30 columns with
unknown variables and each unassigned column must contain a single $1$ in order to intersect
with the first column exactly once.

Once the SAT solver finds a valid~$C$ we programmatically learn the clause
$\bigvee_{p\in C}\lnot p$
which says to block the completion specified by $C$ (in order to perform an exhaustive
search for all completions).  Additionally, suppose that $\varphi$ is a permutation
of~$S$ that fixes the first column (as described above).  We let $\varphi(C)$ denote the set of
true variables in the completion generated by applying $\varphi$
to the completion specified by $C$.
Explicitly, $\varphi(C)$ is computed by applying the column permutations of $\varphi$ to
all the variables in $C$, followed by applying row permutations so that the $1$s
in columns $16$--$21$ remain fixed (as they may have been disturbed by the column
permutations).

For each completion~$C$ found by the SAT solver we record~$C$ and learn the clauses
\[ \bigvee_{p\in\varphi(C)}\lnot p \qquad\text{for all symmetries $\varphi$ fixing the first column} . \]
These clauses block all completions that are equivalent to~$C$, leaving only the nonequivalent
completions in the search space.  We then continue the search in this manner until no completions remain.
At the conclusion of the search we will have recorded a list of all nonequivalent
completions of the upper-left $27\times75$ submatrix of~$P$, i.e., a list of all the completions
up to symmetry.

\begin{figure*}
\centering\tt
\mbox{~~~~~\A}{\color{lightgray}1}{\color{lightgray}0}{\color{lightgray}0}{\color{lightgray}0}{\color{lightgray}0}{\color{lightgray}0}{\color{lightgray}0}{\color{lightgray}0}{\color{lightgray}0}{\color{lightgray}0}{\color{lightgray}0}{\color{lightgray}0}{\color{lightgray}0}{\color{lightgray}0}{\color{lightgray}0}{\color{lightgray}1}{\color{lightgray}0}{\color{lightgray}0}{\color{lightgray}0}{\color{lightgray}0}{\color{lightgray}0}{\color{lightgray}0}{\color{lightgray}0}{\color{lightgray}0}{\color{lightgray}0}{\color{lightgray}0}{\color{lightgray}0}{\color{lightgray}0}{\color{lightgray}0}{\color{lightgray}0}{\color{lightgray}0}{\color{lightgray}0}{\color{lightgray}0}00000{\color{lightgray}0}{\color{lightgray}0}{\color{lightgray}0}{\color{lightgray}0}{\color{lightgray}0}{\color{lightgray}0}0000{\color{lightgray}0}000{\color{lightgray}0}0{\color{lightgray}0}10{\color{lightgray}0}000{\color{lightgray}0}0100{\color{lightgray}0}01100010
\mbox{~~~~~~}{\color{lightgray}1}{\color{lightgray}0}{\color{lightgray}0}{\color{lightgray}0}{\color{lightgray}0}{\color{lightgray}0}{\color{lightgray}0}{\color{lightgray}0}{\color{lightgray}0}{\color{lightgray}0}{\color{lightgray}0}{\color{lightgray}0}{\color{lightgray}0}{\color{lightgray}0}{\color{lightgray}0}{\color{lightgray}0}{\color{lightgray}1}{\color{lightgray}0}{\color{lightgray}0}{\color{lightgray}0}{\color{lightgray}0}{\color{lightgray}0}{\color{lightgray}0}{\color{lightgray}0}{\color{lightgray}0}{\color{lightgray}0}{\color{lightgray}0}{\color{lightgray}0}{\color{lightgray}0}{\color{lightgray}0}{\color{lightgray}0}{\color{lightgray}0}{\color{lightgray}0}00100{\color{lightgray}0}{\color{lightgray}0}{\color{lightgray}0}{\color{lightgray}0}{\color{lightgray}0}{\color{lightgray}0}0010{\color{lightgray}0}101{\color{lightgray}0}0{\color{lightgray}0}00{\color{lightgray}0}000{\color{lightgray}0}0000{\color{lightgray}0}00010000
lines~{\color{lightgray}1}{\color{lightgray}0}{\color{lightgray}0}{\color{lightgray}0}{\color{lightgray}0}{\color{lightgray}0}{\color{lightgray}0}{\color{lightgray}0}{\color{lightgray}0}{\color{lightgray}0}{\color{lightgray}0}{\color{lightgray}0}{\color{lightgray}0}{\color{lightgray}0}{\color{lightgray}0}{\color{lightgray}0}{\color{lightgray}0}{\color{lightgray}1}{\color{lightgray}0}{\color{lightgray}0}{\color{lightgray}0}{\color{lightgray}0}{\color{lightgray}0}{\color{lightgray}0}{\color{lightgray}0}{\color{lightgray}0}{\color{lightgray}0}{\color{lightgray}0}{\color{lightgray}0}{\color{lightgray}0}{\color{lightgray}0}{\color{lightgray}0}{\color{lightgray}0}10000{\color{lightgray}0}{\color{lightgray}0}{\color{lightgray}0}{\color{lightgray}0}{\color{lightgray}0}{\color{lightgray}0}0000{\color{lightgray}0}000{\color{lightgray}0}0{\color{lightgray}0}00{\color{lightgray}0}100{\color{lightgray}0}0010{\color{lightgray}0}10000100
~22-27~{\color{lightgray}1}{\color{lightgray}0}{\color{lightgray}0}{\color{lightgray}0}{\color{lightgray}0}{\color{lightgray}0}{\color{lightgray}0}{\color{lightgray}0}{\color{lightgray}0}{\color{lightgray}0}{\color{lightgray}0}{\color{lightgray}0}{\color{lightgray}0}{\color{lightgray}0}{\color{lightgray}0}{\color{lightgray}0}{\color{lightgray}0}{\color{lightgray}0}{\color{lightgray}1}{\color{lightgray}0}{\color{lightgray}0}{\color{lightgray}0}{\color{lightgray}0}{\color{lightgray}0}{\color{lightgray}0}{\color{lightgray}0}{\color{lightgray}0}{\color{lightgray}0}{\color{lightgray}0}{\color{lightgray}0}{\color{lightgray}0}{\color{lightgray}0}{\color{lightgray}0}01000{\color{lightgray}0}{\color{lightgray}0}{\color{lightgray}0}{\color{lightgray}0}{\color{lightgray}0}{\color{lightgray}0}0100{\color{lightgray}0}000{\color{lightgray}0}0{\color{lightgray}0}00{\color{lightgray}0}001{\color{lightgray}0}1001{\color{lightgray}0}00000000
\mbox{~~~~~}~{\color{lightgray}1}{\color{lightgray}0}{\color{lightgray}0}{\color{lightgray}0}{\color{lightgray}0}{\color{lightgray}0}{\color{lightgray}0}{\color{lightgray}0}{\color{lightgray}0}{\color{lightgray}0}{\color{lightgray}0}{\color{lightgray}0}{\color{lightgray}0}{\color{lightgray}0}{\color{lightgray}0}{\color{lightgray}0}{\color{lightgray}0}{\color{lightgray}0}{\color{lightgray}0}{\color{lightgray}1}{\color{lightgray}0}{\color{lightgray}0}{\color{lightgray}0}{\color{lightgray}0}{\color{lightgray}0}{\color{lightgray}0}{\color{lightgray}0}{\color{lightgray}0}{\color{lightgray}0}{\color{lightgray}0}{\color{lightgray}0}{\color{lightgray}0}{\color{lightgray}0}00001{\color{lightgray}0}{\color{lightgray}0}{\color{lightgray}0}{\color{lightgray}0}{\color{lightgray}0}{\color{lightgray}0}1001{\color{lightgray}0}000{\color{lightgray}0}0{\color{lightgray}0}01{\color{lightgray}0}000{\color{lightgray}0}0000{\color{lightgray}0}00000001
\mbox{~~~~~}\v{\color{lightgray}1}{\color{lightgray}0}{\color{lightgray}0}{\color{lightgray}0}{\color{lightgray}0}{\color{lightgray}0}{\color{lightgray}0}{\color{lightgray}0}{\color{lightgray}0}{\color{lightgray}0}{\color{lightgray}0}{\color{lightgray}0}{\color{lightgray}0}{\color{lightgray}0}{\color{lightgray}0}{\color{lightgray}0}{\color{lightgray}0}{\color{lightgray}0}{\color{lightgray}0}{\color{lightgray}0}{\color{lightgray}1}{\color{lightgray}0}{\color{lightgray}0}{\color{lightgray}0}{\color{lightgray}0}{\color{lightgray}0}{\color{lightgray}0}{\color{lightgray}0}{\color{lightgray}0}{\color{lightgray}0}{\color{lightgray}0}{\color{lightgray}0}{\color{lightgray}0}00010{\color{lightgray}0}{\color{lightgray}0}{\color{lightgray}0}{\color{lightgray}0}{\color{lightgray}0}{\color{lightgray}0}0000{\color{lightgray}0}010{\color{lightgray}0}1{\color{lightgray}0}00{\color{lightgray}0}010{\color{lightgray}0}0000{\color{lightgray}0}00001000
\caption{A completion of rows $22$--$27$ and the first $75$ columns.  Gray entries denote initially known entries and black entries
denote the completions of the initially unassigned entries.}\label{fig:completion}
\end{figure*}

\section{Results}\label{sec:results}

In this section we discuss our implementation, timings, and compare our results with
those of previous searches.

\subsection{Implementation}\label{sec:implementation}

A Python script of less than 200 lines was written to generate a SAT instance containing
the clauses described in Section~\ref{sec:encoding}
(our source code is available at \href{https://uwaterloo.ca/mathcheck/}{\nolinkurl{uwaterloo.ca/mathcheck}}).
The instance contains $51\cdot71=3825$
distinct variables and 79,248 distinct clauses, including 3075 unit clauses and therefore
750 unknown variables.
The symmetry breaking method described in Section~\ref{sec:symbreaking} was implemented
using the programmatic SAT solver MapleSAT~\cite{LiangGPC16} with the symmetry groups
and row and column permutations computed by the computer algebra system Maple 2019~\cite{maple}.

MapleSAT found 42,496 completions of the rows $22$--$27$ of~$P$, of which
1021 of these completions were inequivalent.  This naturally splits the search space
into 1021 distinct subspaces, one for each nonequivalent completion of the rows $22$--$27$.
We now discuss how these completions can be used to help the SAT solver solve
the SAT instance containing the remaining rows $28$--$51$.

The simplest option is simply to generate a distinct SAT instance for each distinct completion
of the rows $22$--$27$ and to include the true variables that appear in the completion as unit clauses.
However, to avoid the overhead of calling a SAT solver 1021 times it is better to generate
a single \emph{incremental SAT} (see~\cite{nadel2012efficient}) instance that contains 1021 sets of assumptions.
A second option is to use a single SAT instance and
add \emph{blocking clauses} for each of the $42{,}496-1021=41{,}475$
completions that are equivalent to one of the remaining $1021$ completions.
For example, if $C$ is the set of variables assigned to true in a completion then
the completion can be blocked by adding the clause $\bigvee_{p\in C}\lnot p$ into the instance.

Experimentally it was determined that it is possible to find completions of~$P$
using the first 45 rows and 75 columns.  In fact, MapleSAT was able to 
find explicit completions of the upper-left $45\times75$ submatrix in about 2 seconds (see Figure~\ref{fig:completion}).
The completion shown in Figure~\ref{fig:completion}
is special in that regardless which representative chosen
for rows $22$--$27$ the representative can always be extended to $45$ rows.  In each of the
other 1020 cases there is at least one representative of the rows $22$--$27$ for which
it is impossible to extend that representative to $45$ rows.

\begin{figure*}
\centering\small\tt
\mbox{~~~~~\A100000000000000100000000000000000001000000000000000100000000001101000000000}
\mbox{~~~~~~100000000000000010000000000000000100000000000100010000000000100000000000100}
\mbox{~~~~~~100000000000000001000000000000000000000000000010000000000000000010001010010}
\mbox{~~~~~~100000000000000000100000000000000010000000000000000000010010000000010100000}
\mbox{~~~~~~100000000000000000010000000000000000010000001001000000001000000000000000001}
\mbox{~~~~~~100000000000000000001000000000000000100000000000001001000001000000000001000}
\mbox{~~~~~~000000000100000000000100000000000000000000000000000001000000100100100000010}
\mbox{~~~~~~000000000100000010000010000000010010000000001000000000000000000000000001000}
\mbox{~~~~~~000000000100000001000001000000001001000000000001000000000010000000000000000}
\mbox{~~~~~~000000000100000000000000100010000000010000000000000100000000000010010000000}
\mbox{~~~~~~000000000100000000100000010100000000100000000000010000000000000000000000001}
\mbox{lines~000000000100000100000000001000000000000000000100100000000001000000001000000}
\mbox{22-45~000000000000001000000000000100000000000000000000000000000000110001001001000}
\mbox{~~~~~~000000000000001000000000010010000100000000000001000001000000000000000100000}
\mbox{~~~~~~000000000000001000000100000001000000000000000000000100000010000000000010001}
\mbox{~~~~~~000000000000001000000000001000100001000000001010000000000000000000010000000}
\mbox{~~~~~~000000000000001000000000000000010000100000000100000000100000001000000000010}
\mbox{~~~~~~000000000000001000000010000000001000010000000000000000010001000000000000100}
\mbox{~~~~~~000000000010000000100000000000010100000000000000000010000000000100000010000}
\mbox{~~~~~~000000000010000100000000000100000010000000000010000000000100000000000000100}
\mbox{~~~~~~000000000010000000000000000000000001000010000000000000011000000000100001000}
\mbox{~~~~~~000000000010000000010000000000000000100000010000100100000000000000000100000}
\mbox{~~~~~~000000000010000001000000000000100000010100000000001000000000100000000000000}
\mbox{~~~~~\v000000000010000000001000000001000000001000000101000000000000000010000000000}
\caption{An assignment to the first $75$ columns and rows $22$--$45$ of $P$ that produces a $45\times75$ partial projective plane.}\label{fig:partial}
\end{figure*}

\subsection{Timings}

We used the SAT solver MapleSAT~\cite{LiangGPC16} running on an Intel i7 CPU
at 2.7 GHz for all timings unless otherwise specified.  The base SAT instance using 51 rows and 75 columns and no
symmetry breaking was shown to be unsatisfiable in 6.3 minutes.
A DRUP proof of size 1.6~GB was produced and
was verified using the proof checker DRAT-trim~\cite{wetzler2014drat}.
Interestingly, the cube-and-conquer paradigm~\cite{heule2017solving}
is particularly effective on SAT instances of this form.
The CnC solver of M.J.H.~Heule~\cite{heule2018cube} outperformed the non-programmatic version of MapleSAT by solving the
SAT instance in 5.2 minutes (1.9 minutes for cubing using March\_cu~\cite{heule2011cube}
and 3.3 minutes for conquering using Glucose~\cite{audemard2009predicting}).
The CnC solver generated a proof of size 1.2 GB (or 410 MB after trimmed and compressed in the binary DRAT format).

Using our programmatic encoding as described in Section~\ref{sec:symbreaking},
MapleSAT found all 42,496 completions (and 1021 nonequivalent completions) of the first 27 rows in 2.4 seconds.
Afterwards, the incremental SAT instance showing that none of the 1021 nonequivalent
completions of the first 27 rows can be extended to 51 rows was solved in 4.4 seconds.
There does not seem to be a standard proof format for incremental SAT instances, so this
run was not formally verified.  However, we did formally verify that the SAT instance
containing the symmetry ``blocking clauses'' was unsatisfiable.  This instance contained
clauses blocking the $41{,}475$ completions of the first 27 rows that are known to be
equivalent to one of the remaining $1021$ completions.  MapleSAT solved this instance
in 13 seconds and produced a DRUP proof of size 147 MB (or 35 MB after trimmed and compressed
in the binary DRAT format, available at \href{https://uwaterloo.ca/mathcheck/}{\nolinkurl{uwaterloo.ca/mathcheck}}).

We did not attempt to optimize the generation of the SAT instances because we
wanted to make the generation scripts as straightforward as possible.
In any case, the generation
of the SAT instances required less than a second using our Python script.

\subsection{Comparison with previous searches}\label{sec:comparison}

There have been a number of previous searches verifying the nonexistence
of projective planes of order ten containing weight 15 codewords~\cite{casiello2010sull,clarkson2014nonexistence,denniston1969non,macwilliams1973existence,perrott2016existence,roy2005proving,roy2011confirmation}.  To the best
of our knowledge, all previous searches have relied on computers in an
essential way---there is no known ``purely theoretical'' explanation for this result.

The first searches by Denniston~\cite{denniston1969non} and MacWilliams, Sloane, and Thompson~\cite{macwilliams1973existence}
used punch cards or paper tape and so we are not able to rerun their searches on modern hardware.
Denniston used~90 minutes on an Elliott 4130 mainframe computer
and MacWilliams et~al.~used 3 hours on
a General Electric 635 mainframe computer.  Their searches were similar though Denniston
used the light rows that intersect the columns numbered 1, 10, 15, and 5,
and some different symmetry removal argumentation.

The next search that we are aware of is by Roy~\cite{roy2005proving} in 2005 and was completed using
a special-purpose search program written in the programming language C, relying on the nauty symbolic
computation library~\cite{mckay2014practical} for solving the graph isomorphism problem.
We ran this program on the same hardware we used for our SAT searches and it completed in
78 minutes.

Casiello, Indaco, and Nagy~\cite{casiello2010sull} completed another search in 2010
using the computer algebra system GAP~\cite{GAP4}.  However, they claim to find no completions
of the first 39 rows and 75 columns (using the same inital 21 rows and 15 columns from
Figure~\ref{fig:matrix}).  Closer inspection of their source code revealed that
incorrect indices were used in one block compatibility check and this caused them to assert
the nonexistence of a partial projective plane that actually exists (as
demonstrated in Figure~\ref{fig:partial}).
After fixing the incorrect indices their program does appear to produce correct results
and requires about 3.3 minutes (on the same hardware from above)
to assert the nonexistence of completions of the first 51 rows and 75 columns.

Next, Clarkson and Whitesides~\cite{clarkson2014nonexistence} completed a search in 2014
using the light rows that intersect the columns numbered 1, 10, 15, 5, and 8.
It was performed using a custom-written and highly optimized search program
in the programming language C.  We obtained the source code from the authors
and it completed in 27 seconds on the same hardware from above.

In 2016, Perrott~\cite{perrott2016existence} completed a search using the computer algebra
system Mathematica~\cite{Mathematica}.  We were not able to run his code on the same
hardware as above (due to the fact Mathematica is proprietary software).
However, his search completed in 55 minutes running on an Intel Xeon X5675 at 3.07~GHz.

Roy~\cite{roy2011confirmation} performed a search in 2011 using a similar approach
as \macwilliams\ except attempting to complete the partial matrix column-by-column
instead of row-by-row and using the columns that were labeled $76$--$111$ in this paper.
These 36 columns can be split into six \emph{blocks} of six columns each, each block
consisting of the columns that are incident to one of the first six rows of~$P$.
This search found no completions of the first four blocks in 2.3 minutes.  We attempted
to verify this result, but found completions of all six blocks
(available at \href{https://uwaterloo.ca/mathcheck/}{\nolinkurl{uwaterloo.ca/mathcheck}}).
Thus, we conclude there was likely an undiscovered bug in the search program of~\cite{roy2011confirmation}
that caused the search to be incomplete.
Unfortunately, this source code is no longer available
so it is impossible to determine the source of the discrepancy for certain.

In comparison, we completed our search using MapleSAT (augmented with
a programmatic learning method as described in Section~\ref{sec:symbreaking}) in 6.8 seconds.
Of this time, 2.4 seconds was used
to exhaustively find all nonequivalent completions of the first 27 rows and
4.4 seconds was used to show that all of these completions do not extend to 51 rows.
Even counting the time it takes to generate the SAT instances and the overhead of
calling MapleSAT, our search terminates in under 8 seconds.  A comparison of the
searches for which source code is available is shown in Table~\ref{tbl:compare}.

\begin{table}
\centering
\begin{tabular}{cccc}
\textbf{Authors} & \textbf{Year} & \textbf{Language} & \textbf{Time} \\
Roy~\cite{roy2005proving} & 2005 & C & 78 minutes \\
Casiello, Indaco, and Nagy~\cite{casiello2010sull} & 2010 & GAP & 3.3 minutes \\
Clarkson and Whitesides~\cite{clarkson2014nonexistence} & 2014 & C & 27 seconds \\
Perrott~\cite{perrott2016existence} & 2016 & Mathematica & 55 minutes \\
Bright et al. [this work] & 2019 & SAT & 6.3 minutes \\
Bright et al. [this work] & 2019 & SAT+CAS & 6.8 seconds
\end{tabular}
\caption{A comparison of the searches for weight 15 codewords for which source code is available.}
\label{tbl:compare}
\end{table}

\section{Related work}\label{sec:relatedwork}

As recounted in the introduction, SAT solvers have been used to
perform searches in many different combinatorial problems.
Some of the first successes were computing
van der Waerden numbers by Kouril and Paul~\shortcite{kouril2008van}
and Ahmed, Kullmann, and Snevily~\shortcite{ahmed2014van},
computing Green--Tao numbers by Kullmann~\shortcite{kullmann2010green},
as well as solving a special case of the Erd\H{o}s discrepancy conjecture
by Konev and Lisitsa~\shortcite{konev2015computer}.
Other more recent combinatorial applications include
proving the Boolean Pythagorean triples conjecture~\cite{heule2016solving} and
a new case of the Ruskey--Savage conjecture~\cite{zulkoski2017combining},
as well as computing Ramsey numbers~\cite{codish2016computing},
Williamson matrices~\cite{bright2018sat},
complex Golay sequences~\cite{bright2018enumeration},
and Schur numbers~\cite{heule2018schur}.

We are not aware of any work prior to ours that used
SAT solvers to search for (or prove the nonexistence of) projective planes.
However, there has been work
formalizing the axioms of projective planes in the theorem prover Coq
by Magaud, Narboux, and Schreck~\shortcite{magaud2008formalizing} and
Braun, Magaud, and Schreck~\shortcite{braun2018formalizing}.

\section{Conclusion}\label{sec:conclusion}

In this paper
we have performed a verification of the first nonexistence
result that was crucial in
the renowned proof that a projective plane of order ten does not exist~\cite{lam1991search}.
In particular, we verified that a projective plane of order ten does not generate codewords
of weight fifteen.  There have been a number of exhaustive searches for such a codeword but
all previous searches are difficult to verify.

In particular, the searches~\cite{denniston1969non,macwilliams1973existence,roy2011confirmation}
rely on code that can no longer be run.
The paper~\cite{casiello2010sull} provides source code
but as described in Section~\ref{sec:comparison} their code has a bug that caused
them to assert the nonexistence of a partial projective plane that we found actually
exists.  The paper~\cite{perrott2016existence}
verifies the same result that we verified in this
paper but does so using about ten pages of sophisticated Mathematica
code, and similarly
the works~\cite{clarkson2014nonexistence,roy2005proving} rely on optimized C programs
that are difficult to verify.

In contrast, we have given a simple translation
of properties of a weight fifteen codeword
into Boolean logic and have shown that these properties
are sufficient to prove that such a codeword cannot exist.
This was done by a simple Python
script that generates a SAT instance
that encodes the necessary properties in
conjunctive normal form.
Furthermore, we solved the resulting SAT instance
and provide a 439 MB formally verifiable certificate that the SAT instance
indeed has no solution (or a 37 MB certificate with symmetry breaking clauses
included).

Additionally, in Section~\ref{sec:symbreaking} we showed how to use a programmatic SAT solver
coupled with a computer algebra system to perform symmetry breaking
and greatly decrease the running time of the search.  In particular,
an off-the-shelf version of MapleSAT completed the search in 6.3 minutes,
while a programmatic version of MapleSAT augmented with our CAS symmetry breaking
method completed the search in 6.8 seconds.  This demonstrates the
utility of the SAT+CAS method and provides further evidence (as originally argued
by~\cite{abraham2015building} and independently by~\cite{zulkoski2015mathcheck}) for the power of the method.

Our work shows for the first time that SAT solvers
can be effectively used in the search for finite projective planes---%
our code is currently the fastest available that can verify the
weight~$15$ nonexistence result (see Table~\ref{tbl:compare}).
Although we do not provide a machine-checkable formal proof
directly from the axioms of a projective plane we have performed
the most rigorous verification of this nonexistence result to date.
In particular, our code is much simpler than the code used in
any previous approach to this problem.  We were able to simplify
the code by relying on a SAT solver to do the hard exhaustive
search work.  As a bonus this also produces a nonexistence
certificate that can be formally independently verified.

We expect the SAT+CAS paradigm to also be useful in proving the
nonexistence of codewords of weight 16 and 19 as well---in fact, our system
has already solved the former case.  The basic SAT encoding
is very similar to the encoding provided in this paper, though the symmetry
breaking method used in this paper is specific to the weight 15 search.
A number of structural differences between the cases cause
the weight 16 and 19 searches to be more complicated---this
will be the subject of future research.

\bibliographystyle{spmpsci}
\bibliography{aaecc}

\end{document}